\pdfoutput=1
\documentclass[10pt]{iopart}
\usepackage{graphicx}
\usepackage{xcolor}
\usepackage{dcolumn}
\usepackage{bm}
\usepackage{amssymb}
\expandafter\let\csname equation*\endcsname\relax
\expandafter\let\csname endequation*\endcsname\relax
\usepackage[fleqn]{amsmath}

\begin{document}

\title[A Nuclear-Based Diagnostic Scheme for Nonthermal Ion Spectrum in HED Plasmas Experiments]{A Nuclear-Based Diagnostic Scheme for Nonthermal Ion Spectrum in High Energy Density Plasmas Experiments}

\author{K. Li$^1$, Z. Y. Liu$^1$, A. D. Liu$^1$ and B. Qiao$^1$}
\address{$^1$ Center for Applied Physics and Technology, HEDPS and State Key Laboratory of Nuclear Physics and Technology, School of Physics, Peking University, Beijing 100871, China}

\ead{bqiao@pku.edu.cn}

\vspace{10pt}
\begin{indented}
\item[]\today
\end{indented}

\begin{abstract}
The nuclear reactions in a plasma system with energy distribution deviated from Maxwellian are proved to have some unique characteristics including those in their product energy spectrum. Based on this, a new nuclear diagnostic scheme for measuring the nonthermal ion energy spectrum in high-energy-density plasmas is proposed, where the effective temperature and spectral peak of the nonthermal ion energy spectrum are uniquely determined by the spectral width and spectral peak of the product energy spectrum. Then, taking the laser-driven magnetic reconnection experiment as an example, according to the particle-in-cell (PIC) simulation results coupled with the proton-boron (pB) fusion reaction, the pB reaction rate is increased by 4 orders of magnitude; meanwhile, the product spectral peak and spectral width are shifted towards higher energy, deviating from the predictions of the thermal equilibrium assumption, both of which are on account of the accelerated and heated suprathermal protons in magnetic reconnection. Furthermore, analyzing a series of simulation results, the new product spectrum analysis approach suitable for the parameter range of laser-driven magnetic reconnection is found.

\end{abstract}

%
\vspace{2pc}

%
%
%

\section{\label{sec:level1}Introduction}

The nonthermal energy spectrum has been proved to exist in a large number of nonthermal equilibrium plasma systems, such as, in astrophysical phenomena, knock-on perturbation in the interior of stars \cite{RN1041}, the magnetic reconnection acceleration at the stellar corona \cite{RN1186,RN1362,RN1363}, and diffusion shock acceleration that occur in the interstellar medium \cite{RN1343}; in the laboratory, the plasma temperature anisotropy in magnetic confinement fusion caused by auxiliary heating \cite{RN1268,RN1272,RN1278}, the Weibel instability occurring in colliding plasmas \cite{RN359}, and magnetic reconnection in the Madison Symmetric Torus (MST) reversed field pinch device \cite{RN183}. Moreover, in laser-driven high energy density (HED) plasmas, the nonthermal energy spectrum is populated by suprathermal fusion products upscattering the reactants \cite{Hayes_2016}, and the magnetic reconnection accelerating and heating the ions \cite{RN1217}. It is proved that the nonthermal energy spectrum is going to significantly affect the energy transport, nuclear reactions \cite{RN1212, RN1060}, and other key physical processes consequently. Therefore, diagnosing the nonthermal distribution of HED plasma is of great significance for understanding the generation of extremely energetic particles in astrophysics and realizing the utilization of fusion energy. However, the nonthermal particles only occupy a very small population causing their signals to be masked by the signals of the thermal particles in the diagnostic data. So far, there is no appropriate approach to measuring the nonthermal ion energy spectrum in the HED experiment.

Over the past decades, the nuclear diagnostic scheme has been proposed \cite{RN2,RN1432} to diagnose the state of HED plasma, which is applied  to diagnose the temperature of fusion fuel in inertial confinement fusion (ICF) \cite{RN1331,RN636,RN1358} and the beam-target reaction in Z-pinch \cite{RN1334} experiments, the magnetization degree of ions in Magneto-ICF (MICF) \cite{RN1257,RN1288} experiments and ion beam parameters in laser-driven ion acceleration experiments. Based on this, several researchers have explored the numerical solution of the product spectrum in different plasma energy spectrum \cite{RN1332,RN1325,RN1326,RN1327,RN1328,RN1357,RN1442}. But so far, there is no research work proposed to diagnose the characteristic physical parameters of nonthermal ion energy spectrum in HED plasma using a nuclear diagnostic scheme.

We propose, for the first time, a nuclear diagnostic scheme to measure the nonthermal ion energy spectrum in HED plasma. Compared with the traditional scheme, which diagnoses directly the emitted ions, this new scheme has two advantages: on the one hand, because the nuclear reaction cross-section strongly depends on particle energy, we can always choose a suitable nuclear reaction to effectively amplify the product signal generated by the nuclear reaction from concerned ion energy range; On the other hand, because most of the nuclear reaction energy release $Q$ is far greater than the plasma temperature, the emitted product signal is not vulnerable to the interference of electromagnetic field and background thermal signal. Therefore, our scheme has a universal guiding significance for experimental diagnosis.

In this paper, the new nuclear diagnostic approach is applied, by analyzing the product spectrum of the nuclear reaction $p(^{11}B,Be^{*})\alpha_0$ (pB), to diagnose the ion nonthermal spectrum in high-energy-density plasmas around several hundred $\rm{keV}$. Firstly, we theoretically deduce the directional energy spectrum of products and correlate the spectral peak and spectral width of the nuclear reaction product spectrum with the peak energy and effective temperature suprathermal ion spectrum. Then, taking the laser-driven magnetic reconnection as an example, we make use of the PIC code including the self-consistently expanded nuclear reaction calculation module \cite{RN1212,RN1355,RN559} to simulate a series of cases and utilize the product spectrum analysis (PSA) approach to diagnose the nonthermal spectrum of protons populated by acceleration and heating effects of the reconnection. We found that fusion events during reconnection occur due to density piling, plasmoid trapping, and particle acceleration, and ultimately increase pB reaction rate by four orders of magnitude. Furthermore, according to the PSA approach, fusion product spectral information can indirectly demonstrate that reconnection accelerates and heats protons and populates a suprathermal tail spectrum. In addition, through the analysis of yield and product spectral information, we give the parameter applicable range of PSA in laser-driven magnetic reconnection.

This paper is organized as follows. In section~\ref{sec:level2}, the brief introduction and verification of theoretical analysis approach of product spectrum. A 2D-PIC simulation of a magnetic reconnection and product spectrum analysis has been done in section~\ref{sec:level3}. Finally, the summary and discussion are given in section~\ref{sec:level4}.

\section{\label{sec:level2}The theoretical principles of product spectrum analysis approach}

\subsection{\label{sec:level2-1}The energy spectrum of nuclear reaction product from nonthermal equilibrium plasma}

In this subsection, referring the derivation of product spectrum from thermal equilibrium ion spectrum as reference \cite{RN1328}, we derive the nuclear reaction product spectrum from a nonthermal ion spectrum. The velocity distribution of ion species $i$ is defined by $F_i(\boldsymbol{v}_i)$, which can be decomposed into a Maxwellian distribution $f_{M}(\boldsymbol{v}_i)$ and multiple suprathermal tail distributions $f_{D,j}(\boldsymbol{v}_i)$ in form of Maxwellian distribution involving high energy peak corresponding to a drift velocity $\boldsymbol{v}_{f,i}$,
\begin{equation}\label{eq1}
F_i(\boldsymbol{v}_i) = (1-\sum_{j}\kappa_j)f_{M}(\boldsymbol{v}_i) + \sum_{j}\kappa_j f_{D,j}(\boldsymbol{v}_i),
\end{equation}
where $0<\kappa_j<1$ is the fraction of suprathermal tail distribution $j$, which only contains a very small number of ion populations, $\sum_{j}\kappa_j \ll 1$. And in this work, for simplicity, the nonthermal distribution only contains one suprathermal tail distribution. The Maxwellian distribution $f_{M}(\boldsymbol{v}_i)$ and suprathermal tail distribution $f_{D}(\boldsymbol{v}_i)$ are defined respectively,
\begin{equation}\label{eq2}
f_M(\boldsymbol{v}_i) = \left(\frac{m_i}{2\pi T_i}\right)^{3/2}\exp{\left(-\frac{m_i v_i^2}{2T_i}\right)},
\end{equation}
and
\begin{equation}\label{eq3}
f_D(\boldsymbol{v}_i) = \left(\frac{m_i}{2\pi T_i}\right)^{3/2}\exp{\left[-\frac{m_i (\boldsymbol{v}_i-\boldsymbol{v}_{f,i})^2}{2T_i}\right]}.
\end{equation}

The number of reaction events per unit time and unit volume between species $1$ with velocity $\boldsymbol{v}_1$ and species $2$ with $\boldsymbol{v}_2$ can be given by 
\begin{equation}\label{eq4}
Y(\boldsymbol{v}_1, \boldsymbol{v}_2) = \frac{n_1 n_2 }{1 + \delta_{12}}v_r \sigma(v_r)F_1(\boldsymbol{v}_1)F_2(\boldsymbol{v}_2)d^3\boldsymbol{v}_1d^3\boldsymbol{v}_2,
\end{equation}
where the $\delta_{12}$ is the Kronecker symbol, $v_r = |\boldsymbol{v}_1-\boldsymbol{v}_2|$, and $\sigma$ is the nuclear reaction cross-section. Because $f_M$ is a simple form of $f_D$ without $\boldsymbol{v}_{f,i}$, we insert the (\ref{eq3}) into (\ref{eq4}) gives universal solution
\begin{align}\label{eq5}
&Y(\boldsymbol{v}_1, \boldsymbol{v}_2) = \Lambda v_r \sigma(v_r)\exp{\left[-\frac{m_1(\boldsymbol{v}_1-\boldsymbol{v}_{f,1})^2}{2T_1}\right.}\nonumber \\
& {\left.-\frac{m_2(\boldsymbol{v}_2-\boldsymbol{v}_{f,2})^2}{2T_2}\right]}d^3\boldsymbol{v}_1d^3\boldsymbol{v}_2,
\end{align}
where
\begin{equation}\label{eq6}
\Lambda = \frac{n_1n_2}{1+ \delta_{12}}\left(\frac{m_1m_2}{4\pi^2 T_1 T_2}\right)^{3/2}.
\end{equation}
We transform the particle velocity into relative velocity $\boldsymbol{v}_r$ and center-of-mass~(COM) velocity $\boldsymbol{v}_c$
\begin{equation}\label{eq7}
\begin{cases}
\boldsymbol{v}_1 = \boldsymbol{v}_c + \frac{\mu}{m_1}\boldsymbol{v}_r\\
\boldsymbol{v}_2 = \boldsymbol{v}_c - \frac{\mu}{m_2}\boldsymbol{v}_r
\end{cases},
\end{equation}
and the same to drift velocity
\begin{equation}\label{eq8}
\begin{cases}
\boldsymbol{v}_{f,1} = \boldsymbol{v}_{fc} + \frac{\mu}{m_1}\boldsymbol{v}_{fr}\\
\boldsymbol{v}_{f,2} = \boldsymbol{v}_{fc} - \frac{\mu}{m_2}\boldsymbol{v}_{fr}
\end{cases}.
\end{equation}
Following the Jacobian determinant, the differential is given by  $d^3\boldsymbol{v}_1d^3\boldsymbol{v}_2 \rightarrow d^3\boldsymbol{v}_rd^3\boldsymbol{v}_c$. This result in
\begin{align}\label{eq9}
&Y(\boldsymbol{v}_r, \boldsymbol{v}_c) = \Lambda v_r \sigma(v_r)\exp\left[-\alpha v_c^2+\alpha \boldsymbol{v}_c \cdot \boldsymbol{v}_{fc} - \right.\nonumber\\
&\left.\left(\frac{m_1}{2T_1}v^2_{f,1}+\frac{m_2}{2T_2}v^2_{f,2}\right)-\beta v_r^2 +\beta \boldsymbol{v}_r \cdot \boldsymbol{v}_{fr} \right]d^3\boldsymbol{v}_rd^3\boldsymbol{v}_c.
\end{align}
For simplicity, the temperature of two reactants are assumed to be close $T_1 \approx T_2 =T$, 
\begin{equation}\label{eq10}
\begin{cases}
\alpha = \frac{m_1}{2T_1} + \frac{m_2}{2T_2} \sim \frac{m_1 + m_2}{2T}\\
\beta=\frac{\mu}{m_1 + m_2}\left(\frac{m_2}{2T_1} + \frac{m_1}{2T_2}\right)\sim\frac{\mu}{2T}\\ \gamma=\frac{1}{T_1} - \frac{1}{T_2} \sim 0
\end{cases}.
\end{equation}
For nonthermal distribution, the differential of relative velocity in spherical coordinate is $d^3\boldsymbol{v}_r = v_r^2\sin \theta dv_r d\theta_r d\phi_r$, where the $\theta_r$ and $\phi_r$ are the polar angle and azimuthal angle of relative velocity in the motion coordinate system with velocity $\boldsymbol{v}_{fr}$. Then we integrate the terms including $\theta_r$ and $\phi_r$,
\begin{align}\label{eq11}
&\int_0^{\pi}\int_0^{2\pi}\sin\theta_r\exp[\beta v_rv_{fr}\cos\theta_r]d\phi_rd\theta_r \nonumber\\
&= 4\pi\frac{\sinh(\beta v_rv_{fr})}{\beta v_rv_{fr}},
\end{align}
which gives
\begin{align}\label{eq12}
&Y(\boldsymbol{v}_c, v_r) = \frac{4\pi\Lambda}{\beta v_{fr}} v^2_r \sigma(v_r)\exp\left[ - \left(\frac{m_1}{2T_1}v^2_{f,1}+\frac{m_2}{2T_2}v^2_{f,2}\right)\right.\nonumber\\
&\left.-\alpha v_c^2-\beta v_r^2+\alpha \boldsymbol{v}_c \cdot \boldsymbol{v}_{fc}\right]\sinh(\beta v_rv_{fr}) dv_rd^3\boldsymbol{v}_c.
\end{align}
In addition, the kinetic energy of product species $3$ in COM frame follows the energy conservation and momentum conservation equations giving
\begin{equation}\label{eq13}
\frac{1}{2}m_3u_3^2 = \frac{m_3}{\eta}\left(Q + \frac{1}{2}\mu v_r^2\right),
\end{equation}
where $\eta = m_3(m_3 + m_4)/m_4$, and the transformation between $v_r$ and the velocity of species $3$ in the COM frame $u_3$ follows the Jacobian determinant
\begin{equation}\label{eq14}
dv_r = \frac{u_3 \eta}{\mu v_r}du_3.
\end{equation}
Defining $\zeta=\sqrt{2(\eta u_3^2/2-Q)/\mu}$, the equation (\ref{eq12}) is rewritten as
\begin{align}\label{eq15}
&Y(u_3, \boldsymbol{v}_c) = \frac{4\pi \eta \Lambda}{\beta \mu v_{fr}}u_3 \zeta\sigma(\zeta)\exp(-\beta v_r^2)\sinh(\beta v_rv_{fr})\nonumber\\
&\times\exp\left[-\left(\frac{m_1}{2T_1}v^2_{f,1}+\frac{m_2}{2T_2}v^2_{f,2}\right)\right]\nonumber\\
&\times\exp\left[-\alpha( v_c^2-\boldsymbol{v}_c \cdot \boldsymbol{v}_{fc})\right] du_3 d^3 \boldsymbol{v}_c.
\end{align}
As Galileo transformation $\boldsymbol{v}_c = \boldsymbol{v}_3 - \boldsymbol{u}_3$, and the reaction cross-section is independent on the particle emission angle, which simply requires that we multiply (\ref{eq15}) by a factor $\sin \theta_3(4\pi)^{-1}d\phi_3d\theta_3$ to obtain $Y(\boldsymbol{u}_3, \boldsymbol{v}_c)$ with $\boldsymbol{u}_3$ defined in spherical coordinates, where the $\theta_3$ and $\phi_3$ are the polar angle and azimuthal angle of $u_3$ in COM frame. The yield is
\begin{align}\label{eq16}
&Y(\boldsymbol{u}_3, \boldsymbol{v}_c) = \frac{\eta \Lambda}{\beta \mu v_{fr}}u_3 \zeta\sigma(\zeta)\exp(-\beta v_r^2)\sinh(\beta v_rv_{fr})\nonumber\\
&\times\exp\left[-\left(\frac{m_1}{2T_1}v^2_{f,1}+\frac{m_2}{2T_2}v^2_{f,2}\right)\right]\nonumber\\
&\times\exp\left[-\alpha( \xi-\boldsymbol{v}_c \cdot \boldsymbol{v}_{fc})\right]\sin \theta_3 du_3d\theta_3d\phi_3 d^3 \boldsymbol{v}_3,
\end{align}
where $\xi = (\boldsymbol{v}_3-\boldsymbol{u}_3)^2$. The same to the derivation of equation (\ref{eq11}), we integrate the terms including $\theta_3$ and $\phi_3$,
\begin{align}\label{eq17}
&Y(\boldsymbol{v}_3, \boldsymbol{u}_3) = \frac{4\pi \Lambda \eta}{\mu \alpha \beta v_{fr}|\boldsymbol{v}_{fc}-2\boldsymbol{v}_{3}|}\zeta\sigma(\zeta)\nonumber\\
&\times\exp{\left[ -\left( \frac{m_1}{2T}v^2_{f,1} + \frac{m_2}{2T}v^2_{f,2} \right) \right]}\exp(-\beta\zeta^2)\sinh(\beta \zeta v_{fr})\nonumber\\
&\times\exp\left[-\alpha(v_3^2 + u_3^2 - \boldsymbol{v}_3\cdot \boldsymbol{u}_3)\right]\sinh( \alpha u_3 |\boldsymbol{v}_{fc}-2\boldsymbol{v}_{3}|)\nonumber \\
&\times du_3d^3\boldsymbol{v}_3.
\end{align}
Finally, we replace the $v_3$ with $E_3$, according to $d^3\boldsymbol{v}_3 = v_3^2dv_3d\Omega=v_3/m_3 dE_3d\Omega$ in equation (\ref{eq17}), where $\Omega$ is solid angle. We concern the direction of acceleration, and the Gamow peak is usually much smaller than the fusion energy $Q$, so the detected product spectrum parallel to the direction of acceleration is simplified as
\begin{align}\label{eq18}
&\frac{d^2Y}{dE_3d\Omega}\sim\int \sigma(E_r)\exp\left(-\frac{E_r}{T}\right)\sinh\left(\frac{v_{fr} \sqrt{2\mu E_r}}{2T}\right)\nonumber \\
&\times \exp\left[-\frac{2\alpha}{m_3}\left(\sqrt{E_3} - \sqrt{m_3(E_r + Q)/\eta}\right.\right. \nonumber \\
&\left.\left.\mp \sqrt{m_3/2}\ v_{fc}/2\right)^2\right]dE_r.
\end{align}
Furthermore, for simplicity, the product spectrum can be replaced approximately by the Gaussian distribution \cite{RN1328}, with the same spectral peak $E_0$ and spectral width $\sigma_0$,
\begin{align}\label{eq19}
&\frac{d^2Y}{dE_3d\Omega}\sim\int \sigma(E_r)\exp\left(-\frac{E_r}{T}\right)\sinh\left(\frac{v_{fr} \sqrt{2\mu E_r}}{2T}\right)dE_r \nonumber\\ &\times\exp\left[\frac{(E_3-E_0)^2}{2\sigma_0^2}\right],
\end{align}
where
\begin{equation}\label{eq20}
\begin{cases}
E_0 = \left(\sqrt{\frac{m_3(E_r+Q)}{\eta}} \pm \sqrt{\frac{m_3}{8}}v_{fc} \right)^2\\
\sigma_0 = m_3(\frac{\sqrt{\ln2/2}}{2\alpha}+\sqrt{\frac{E_0}{\alpha m_3}})
\end{cases}.
\end{equation}
The relationship (\ref{eq20}) relates the spectral peak and spectral width of product spectrum with drift velocity and effective temperature of suprathermal ion spectrum. If only reactant species $1$ has suprathermal tail distribution, the (\ref{eq20}) can be rewritten as
\begin{equation}\label{eq21}
\begin{cases}
E_0 = \left(\sqrt{\frac{m_3(E_r+Q)}{\eta}} \pm \sqrt{\frac{m_3}{8}}\frac{\mu}{m_2}v_{f,1} \right)^2\\
\sigma_0 = m_3(\frac{\sqrt{\ln2/2}}{2\alpha}+\sqrt{\frac{E_0}{\alpha m_3}})
\end{cases}.
\end{equation}
If the plasma is thermal equilibrium, the relationship (\ref{eq20}) degenertates to
\begin{equation}\label{eq22}
\begin{cases}
E_0 = \frac{m_3(E_r+Q)}{\eta} \\
\sigma_0 = m_3(\frac{\sqrt{\ln2/2}}{2\alpha}+\sqrt{\frac{E_0}{\alpha m_3}})
\end{cases}.
\end{equation}

Then we can solve the effective temperature $T$ and the peak energy $E_f = m_1 v_{f,1}^2/2$ of suprathermal tail spectrum, according to (\ref{eq20}-\ref{eq22}). In order to more intuitively study the relationship between $E_0$, $\sigma_0$ of the product spectrum with $E_f$, $T$ of the suprathermal tail spectrum, we have defined the spectral moments as the combination of the spectral peak shift of product spectrum and effective temperature of suprathermal tail spectrum $(\Delta E, T)$, where $\Delta E = E_0 - m_3 Q/\eta$ \cite{RN1442}. For thermal equilibrium plasma with certain nuclear reaction, the detected spectral moments should be located in a certain curve in $\Delta E-T$ space in any azimuth. Therefore, the comparison between (\ref{eq20}) and (\ref{eq22}) can obtain the peak energy and effective temperature information of suprathermal tail distribution.

\subsection{\label{sec:level2-2}Verification by PIC simulation}
In this subsection, we make use of the product spectrum of pB reaction to derive the peak energy and effective temperature of the suprathermal spectrum. Only the protons are assumed to have high energy spectrum structure, but the Borons keep thermal equilibrium with the same temperature with protons. We are going to investigate the accuracy of diagnostic theory in variable effective temperature and peak energy of suprathermal protons as follow.

To simulate the nuclear reaction occurs in plasma, a self-consistently extended nuclear reaction calcualtion module, the two-body reactions between macro-particles within each cell are considered in the PIC code, which yields alpha particles with weight \cite{RN1355},
\begin{equation}\label{eq23}
W_\alpha = W_p W_{^{11}B} v_r \sigma \Delta t / V,
\end{equation}
where $W$ is the particle weight, $\Delta t$, and $V$ are the time step and the cell volume, respectively. The calculation of equation (\ref{eq22}) is coupled into PIC code using the Monte-Carlo sampling method with a random number, where each sampling satisfies Poisson distribution with definite expectation equal to $W_\alpha$.

\begin{figure*}[htbp]
	\centering
	\includegraphics[width=0.95\textwidth]{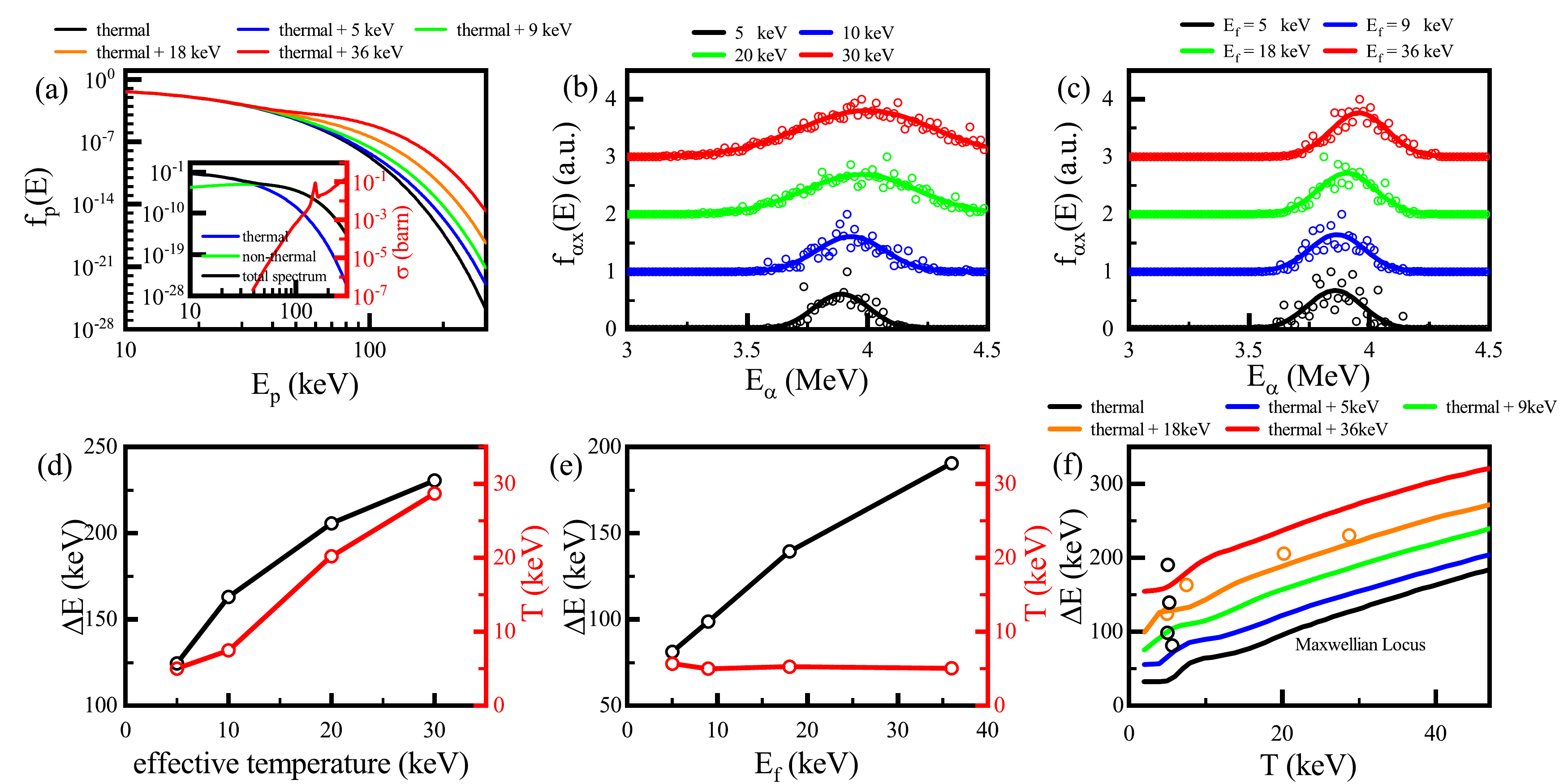}
	\caption{(color online). Theoretical verification by comparing numerical calculation with PIC simulations. (a) The non-Maxwellian distribution with suprathermal tail distribution comparing with nuclear reaction cross-section \cite{RN1006}. (b) and (c) The detected spectrum of $\alpha$ particles along the x-axis within $5^\circ$ varying temperature and peak energy, respectively. (d) and (e) The derived spectral peak shift and effective temperature from the Gaussian distribution varying temperature and peak energy, respectively. (f) The spectral moments space.}
	\label{fig1}
\end{figure*}

We assume a simple toy model, such as equation (\ref{eq1}), where the proton energy spectrum consists of a thermal equilibrium distribution and a suprathermal tail distribution in the x-direction with peak energy $E_f = m_p v_f^2/2$, where $\kappa = 0.05$, as shown in the inset in figure \ref{fig1}(a). As we all know, thanks to the resonance peak energy ($675\ \rm{keV}$) of the pB reaction cross-section, the contribution of the thermonuclear reaction is much smaller than the yield from the high-energy protons, so the $\alpha$ particles are mainly produced by the interaction between suprathermal protons with Borons.

Therefore, we detect the alpha particle energy spectrum within $5^\circ$ along the x-axis from the simulation results. As shown in figure \ref{fig1}(b), when the peak energy is fixed and the plasma temperature increases, the spectral peak of the $\alpha$ particle spectrum shifts to the right, and the spectral width broadens. And it is revealed in (d) that the increase of the spectral peak is proportional to $\ln{T}$, and the calculated effective temperature is consistent with the plasma temperature. On the other hand, when the temperature is fixed, the drift energy increases, and the center of the spectrum shown in (c) also shifts to the right, while the spectral width remains unchanged. And (e) shows that the right shift of the spectral center is positively correlated with the drift energy, and the calculated temperature is still highly consistent with the effective temperature. Moreover, the spectral moments data obtained from a large number of simulations are in good agreement with the numerical results, as shown in (f). It is not difficult to understand that the more the spectral moment deviates from the product spectrum of the Maxwellian Locus, the more significant the nonthermal structure of the ion energy spectrum is. In addition, the variable drift energy modifies spectral moments vertically in $\Delta E - T$ space, while the changeable temperature moves it along the curves shown in (f), which helps diagnose the suprathermal tail of the ion in the HED experiment.

\section{\label{sec:level3}The product spectrum analysis approach for laser-driven magnetic reconnection}
\subsection{\label{sec:level3-1}The simulation settings of magnetic reconnection}

\begin{figure*}[htbp]
	\centering
	\includegraphics[width=0.95\textwidth]{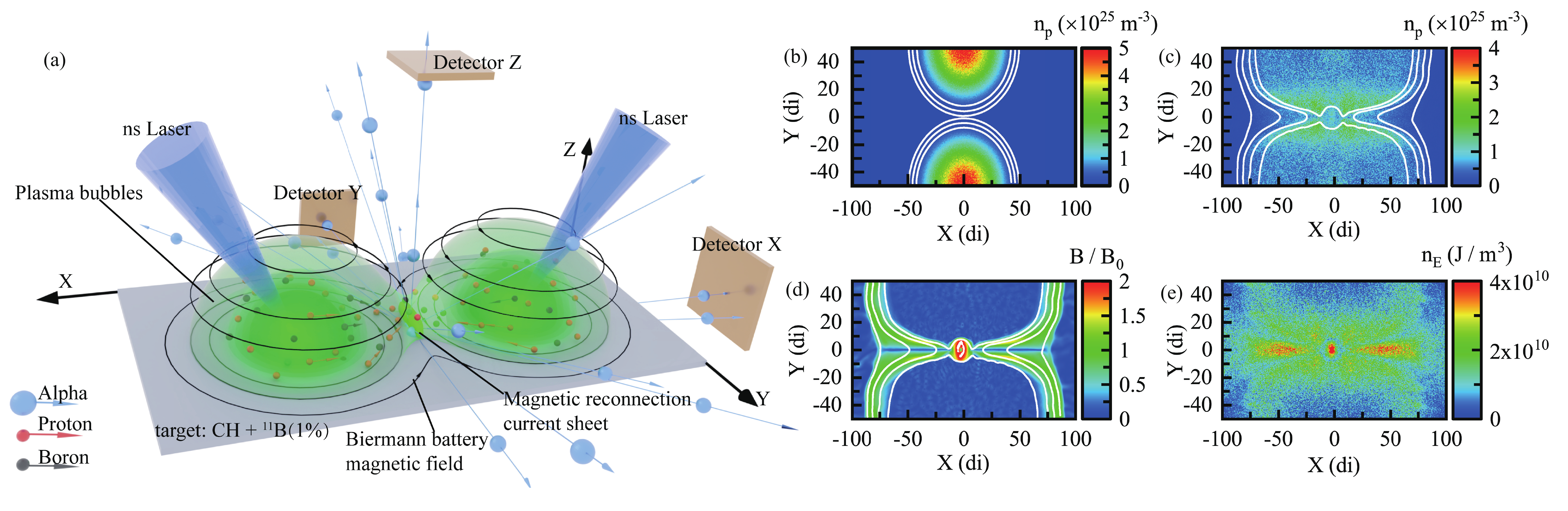}
	\caption{(color online). The schematic of the design of the laser-driven magnetic reconnection for diagnosing suprathermal protons by using pB reaction and the simulation settings and the field distribution at the moment of reconnection. (a) The schematic of the simulation setting. The color transparency indicates the density of bubbles. (b) The initial proton density and magnetic force lines. (c) At $t=0.4 t_d$, the proton density and magnetic force lines. (d) The normalized magnetic strength distribution at $0.4t_d$. (e) The energy density distribution of protons at $0.4 t_d$}
	\label{fig2}
\end{figure*}

To demonstrate the experimental practicality of the product spectrum analysis approach, PSA is used to diagnose accelerated high-energy protons in laser-driven magnetic reconnection. Based on the experimental scheme of Nilson \cite{RN1217}, we have improved the experimental plan by mixing Boron-11, which can maintain the physical process of magnetic reconnection and also couple the energy-sensitive nuclear reaction. The yields and energy spectrum of products from the pB reaction can reflect the states of suprathermal ions in the reconnection current sheet (CS). As our design, figure \ref{fig2}(a) shows that two closely focusing heater beams irradiate a CH planar target foil mixed 1\% Boron, producing two expanding hemispherical bubbles, which carry with it a frozen-in magnetic field formed by the Biermann battery mechanism ($\nabla n_e \times \nabla T_e$). When the two expanding plasma eventually encounters each other and continuously compress, the opposing magnetic field lines break up within the neutral CS and reconnect at downstream of outflow. The energy released as the magnetic field rearrangement drives plasma flows, heats the plasma, and accelerates particles, which can stimulate the nuclear reaction events and produce energetic $\alpha_0$ particles. In addition, three detectors, typically Thomson parabola specrtometer (TPS) in experiments, are placed in the extension line of the CS (Detector X), the in-plane (Detector Y), and out-plane (Detector Z) perpendicular bisectors of CS, respectively to detect the energy spectrum of $\alpha_0$ particles within the same solid angle.

Similar to those previously presented in Refs. \cite{RN1216,RN1221}, A series of two-dimensional (2D) PIC simulations with a newly-developed nuclear reaction calculation module are carried out. Our simulations adopt an initial condition corresponding to a time about halfway through the experiments, after the bubbles are created, have expanded, and have generated their magnetic ribbons, but before the pair of bubbles begin to interact, as shown in figure \ref{fig2}(b). Within the 2D rectangular domain, the centers of the bubbles are given by the vectors $\boldsymbol{R}^{(1)} = (0, L_n)$ and $\boldsymbol{R}^{(2)} = (0, -L_n)$, where $L_n$ is the radius of each bubble when they begin to interact, and the corresponding radial from the the bubble centers are $\boldsymbol{r}^{(i)} = \boldsymbol{r} - \boldsymbol{R}^{(i)}$. The initial density profile is given by $n_b + n^{(1)} + n^{(2)}$, where $n_b$ and $n_0$ are respectively the background density and the initial maximum density, and the densities $n^{(i)}$ are
\begin{equation}\label{eq24}
n^{(i)}=(n_0 - n_b)\cos^2(\pi r^{(i)} / 2L_n)H(L_n - r^{(i)}).
\end{equation}
The expansion velocity is given by $\boldsymbol{V} = \boldsymbol{V}^{(1)} + \boldsymbol{V}^{(2)}$, where the $\boldsymbol{V}^{(i)}$ are
\begin{equation}\label{eq25}
\boldsymbol{V}^{(i)} = V_0\sin(\pi r^{(i)} /L_n) \boldsymbol{\hat{r}}^{(i)} H(L_n - r^{(i)}).
\end{equation}
The initial magnetic field corresponds to the sum of two oppositely aligned ribbons of finite flux, $\boldsymbol{B} = \boldsymbol{B}^{(1)} + \boldsymbol{B}^{(2)}$, where 
\begin{align}\label{eq26}
\boldsymbol{B}^{(i)} = &B_0\sin[\pi(L_n-r^{(i)})/2L_b]\boldsymbol{\hat{r}}^{(i)}\times \boldsymbol{\hat{z}}\nonumber \\
&\cdot H\{[r^{(i)}-(L_n - 2L_b)](L_n - r^{(i)})\}.
\end{align}
Here $L_B = L_n/4$ is the initial half-width of the magnetic field ribbon. $n_0$, $V_0$ and $B_0$ are initial parameters, and $H(x)$ is the Heaviside step fuction. In addition, the initial electric field is $\boldsymbol{E} = -\boldsymbol{V}\times\boldsymbol{B}/c$, consistent with the initial motion of the magnetized plasma. An initial out-of-plane current is included that is consistent with $\nabla \times \boldsymbol{B} = 4\pi \boldsymbol{J} / c$ and distributed to the electrons and ions by the inverse of their mass ratio.

The temperature is initially uniform as $T_e = T_i = T_0 = 0.01 m_e c^2$, the plasma beta $\beta_e = 2(M_A/M_S)^2 = 4 - 32$, where $M_A=V_0/V_A$ and $M_S=V_0/C_S = 0.5-3$ are Alfvenic and sonic Mach numbers, the Alfven speed $V_A = B_0/\sqrt{4\pi n_0 m_i}$ and sound speed $C_S = \sqrt{ZT_e/m_i}$. Define the dimensionless parameter $S = 2L_n/d_i = 20 - 100$ corresponding to the scales of the reconnection CS, where the $d_i$ is the skin depth of ions. For the numerical parameters of the simulations, the length of the domain is given by $4L_n \times 2L_n$ in the $(x, y)$ coordinates, and the resolution of the spatial grid is $\Delta x = 0.33d_e \approx 0.033d_i$. The total simulation time is the dynamical time $t_d = L_n / C_S$, and the discrete time step is $\Delta t = 0.33\omega_{pe}^{-1}$.

\begin{figure*}[b]
	\centering
	\includegraphics[width=0.95\textwidth]{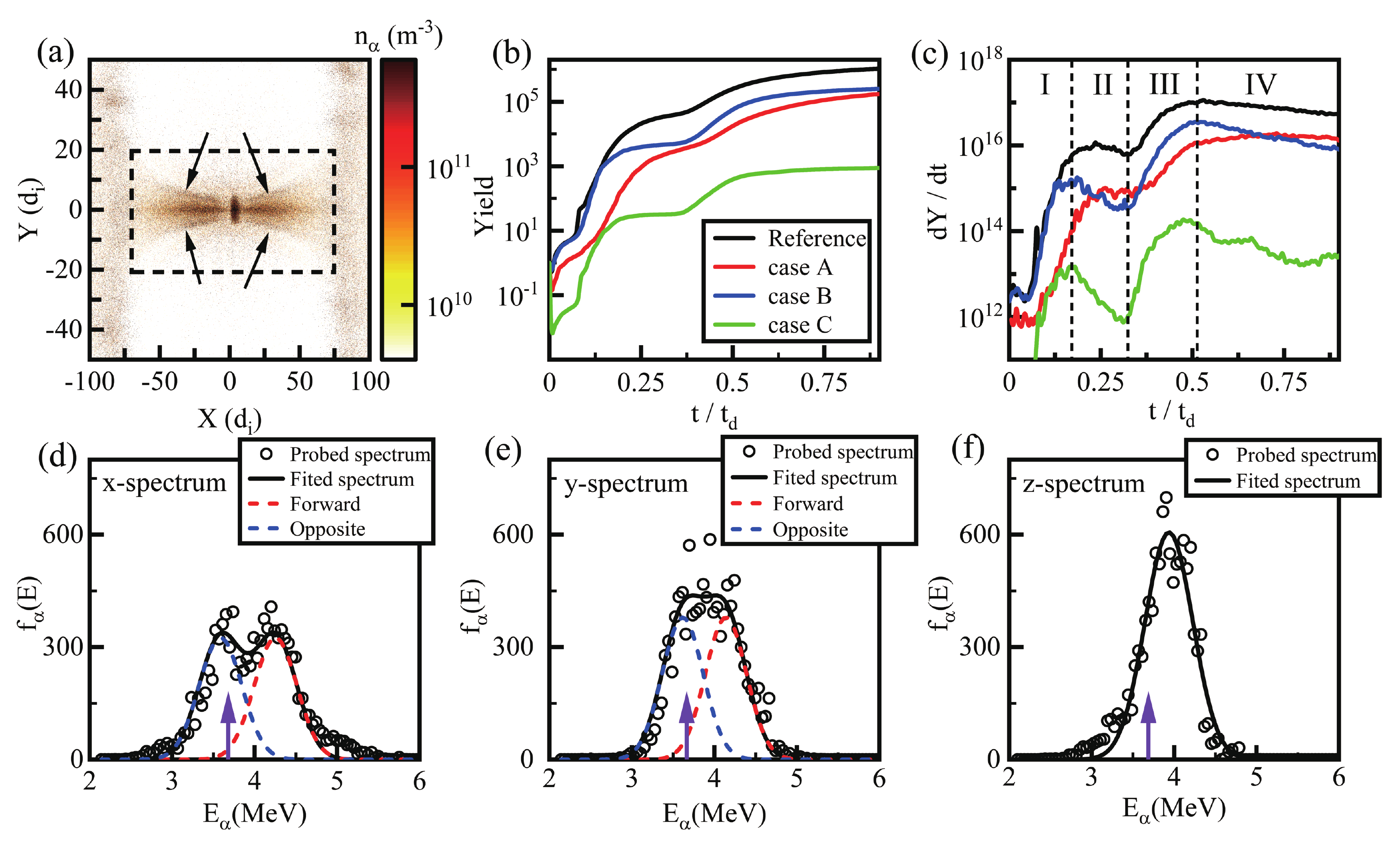}
	\caption{(color). The yields and energy spectrum of products. (a) The reaction position. (b) The yields evolution with time at different parameters. The Reference case involves $\beta = 8$, $M_s = 2$ and $S=100$. The cases A, B and C indicate $\beta = 16$, $M_s = 1$ and $S=50$, respectively maintaining other parameters. (c) The reaction rates evolution with time at different parameters. (d), (e) and (f), The alpha particles energy spectrums detected by X, Y and Z detector, respectively. }
	\label{fig3}
\end{figure*}

To reduce the demand for computational resources down to the practical extent, the ion-to-electron mass ratio has been reduced by parameter $f_m$ between the scaled~(with subscript '1') and the real~(with subscript '0') systems, with $m_p = 100m_e$. This is a commonly used approximation and can still allow for adequate separation of the relevant temporal and spatial scales to accurately model the physical processes \cite{RN934}. Meanwhile, the temperature has been scaled by $f_T = 25$. To remain the Vlasov equation, the self-similar temporal and spatial scales and electromagnetic fields can be easily set up, where the key parameters $\beta$, $S$, and $M_s$ remain the same. Furthermore, because the nuclear reaction cross-section is the nonlinear function of energy, the energy in simulation needs to be scaled up to the natural energy in each step of simulation.

As shown in figure \ref{fig2}(c), the two expanding bubbles encounter each other in the central region at $ t = 0.4 t_d $. The frozen magnetic field prevents the interpenetration of the inflow which piles the plasma up forming the CS with the density increased from $ 0.1n_0 $ to $2n_0$ upstream. Meanwhile, the magnetic field in (d) has been enhanced up to 2 times the initial $ B_0 $, and a plasmoid form in the CS when the magnetic field lines rearrange. According to the generalized Ohm’s law, inside the CS, the out-of-plane electric field is predominantly supported by the Hall and off-diagonal electron pressure tensor terms \cite{RN1216}, which causes an opposite electric field in the CS and the plasmoid; outside the CS, the Hall term lead by electron-proton decoupling supports the in-plane electric field known as in-plane Hall electric field, which produces about $ 1\ \rm{MV} $ potential difference between upstream and downstream. Because of the combination of the pile of the proton density, the direct acceleration by reconnection electric field, and Hall electric field acceleration, the protons can obtain significant energy density in the plasmoid and downstream, as shown in (e).

\subsection{\label{sec:level3-2}The product spectrum analysis for magnetic reconnection}

According to, 
\begin{equation}\label{eq27}
Y(\boldsymbol{r}) = \int n_p(\boldsymbol{r},t)n_{^{11}B}(\boldsymbol{r},t)\langle\sigma v\rangle(\boldsymbol{r},t) dt,
\end{equation}
the initial spatial distribution of the products depends on the densities of the reactants and the local reactivity. As mentioned in the previous subsection, the reconnection process will trigger a large number of local nuclear reaction events due to density piling, plasmoid trapping, and outflow particle acceleration. In the square area as shown in figure \ref{fig3}(a), it occupies the majority of the total yield and has an obvious spatial structure. Firstly, when two bubbles encounter and squeeze each other, a neutral current sheet is formed, which will cause the central plasma temperature to increase and the density to accumulate, so a large number of alpha particles are generated at the upstream boundary indicated by the arrows in (a); then, part of the inflowing protons in the CS are trapped by the plasmoid and accelerated by the z-direction reconnection electric field \cite{RN558}, and undergo a large number of nuclear reactions with the background Boron particles; finally, the inflowing ions are deflected and accelerated to the downstream due to the in-plane Hall electric field \cite{RN184} and participate in pB reaction. In addition, the products outside the square are mainly from the reaction between the expanding bubbles and the low-density background plasma, which have negligible influence on the product spectrum.

On the other hand, the nuclear reaction yields over time are compared for different parameters, as shown in figure \ref{fig3}(b). Firstly, the nuclear reaction yields increase in a stepwise manner, and the total yield can reach the order of $10^6$, which is sufficient to be measured experimentally. Secondly, according to the inference of the maximum energy after the acceleration of electrons and ions by the reconnection electric field mentioned in the literature \cite{RN1221}, $E_{max} \propto M_s S/\beta$ can be obtained. Therefore, compared with the reference case, the $\beta$ of case A is 16, and the magnetic field gradient caused by the accumulation of plasma is smaller, so the weakened reconnection electric field accelerates and heats protons to lower energy, and finally, the total yield can only reach $10^5$. In case B, $M_s=1$, the proton inflow velocity becomes slower, and a large number of them cannot be trapped and heated by the plasmoid, so the increase of its yield is relatively low, which only reaches $10^5$. Moreover, the $S=50$ of case C, due to the reduction of density piling, the shortening of acceleration distance, and the reduction of nuclear reaction time and space, the final nuclear reaction yield is reduced by three orders of magnitude. It can be seen whether the product spectrum analysis approach works strongly dependent on the size of the reconnection current sheet.

In addition, the nuclear reaction rate differs significantly at different stages of reconnection, as in figure \ref{fig3}(c). We divide the evolution of the nuclear reaction rate into four stages. Stage I, when the bubbles just encounter, has an initial reaction rate at a low level ($10^{12}\ \rm{s^{-1}}$) because the frozen magnetic field prevents their interpenetration. Then, the bubbles squeeze each other, pile up the plasma density, and heat plasma, leading to an increase in the nuclear reaction rate by 3 orders of magnitude. Stage II, when reconnection has not yet occurred, but density piling has reached its maximum, the first plateau structure appears. Stage III, when the reconnection rate reaches its maximum, the protons gain energy from the magnetic field in a short time and undergo nuclear reactions with borons within the current sheet, leading to another order of magnitude increase. At stage IV, when the reconnection ends, the dilution of density and the thermalization of the energetic ions together leads to a slight decrease in the reaction rate, and a second plateau structure emerges. Overall, the nuclear reaction rate at the end of the reconnection rises by four orders of magnitude over that at the beginning of the reconnection. And because of the highest yield rate of phase III and the accumulation of phase IV, they mainly contribute to the total yield measured in the detector.

Considering the product spectrum analysis method mentioned in the previous section, the Gaussian distribution has been used to fit the energy spectrum of alpha particles detected in the x-, y-, and z-axes within $5^\circ$, as shown in figures \ref{fig3}(d-f). Parallel to the x-axis, the protons are accelerated symmetrically in the forward and reverse directions of the x-axis, so the detected x-spectrum should be fitted by two Gaussian distributions with similar heights and widths, as shown in (d). Similarly, parallel to the y-axis, protons flow into the reconnection region symmetrically, so the y-spectrum should also be fitted by two Gaussian distributions, as shown in (e). However, parallel to the z-axis, protons can only be accelerated by an out-of-plane electric field along the positive z-axis, so the z-spectrum is fitted by only one Gaussian distribution, as shown in (f).

\begin{figure*}[htbp]
	\centering
	\includegraphics[width=0.95\textwidth]{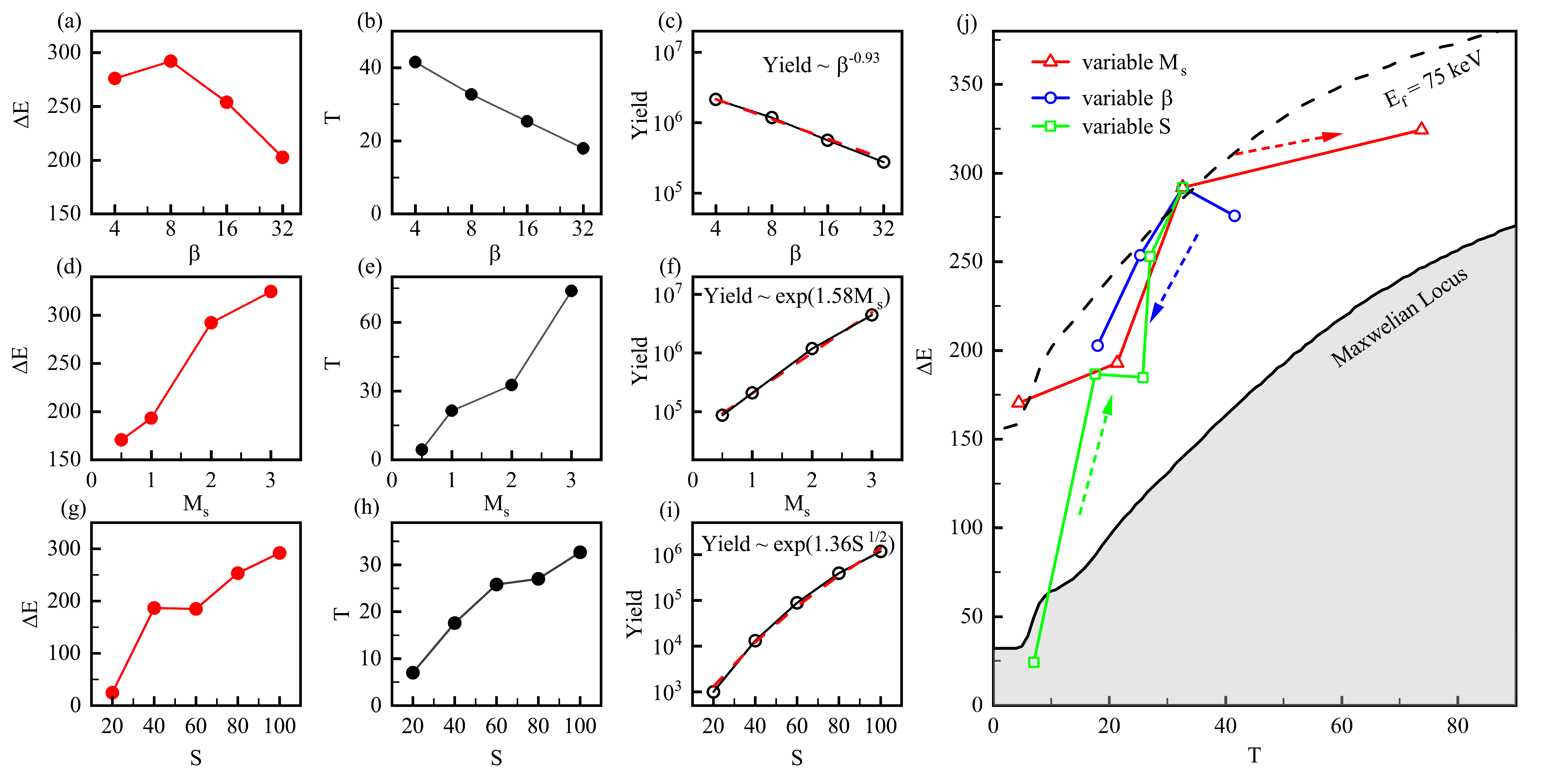}
	\caption{(color). The analysis of yields and spectral moments. At variable $\beta$, inflow velocity $M_s=V_{in} / Vs$ and scales of CS $S = 2L_n/d_i$ respectively in three columns, (a) (d) (g) the spectral center shift, (b) (e) (h) the spectral temperature, (c) (f) (i) the total yields. (j) The diagnosed spectral moments of cases comparing with Maxwellian Locus and non-Maxwellian curve with $E_f = 75\ \rm{keV}$. The arrows indicate the direction of increasing parameters.}
	\label{fig4}
\end{figure*}

As shown in figures \ref{fig4}, a series of simulations have been carried out to quantitatively analyze the evolution of the spectral moments with the reconnection parameters. The three rows of subplots represent the initial $ \beta $ value, the acoustic Mach number $ M_s $, and the reconnection scale $ S $; the three columns of subplots show the spectral center shift, spectral temperature, and the total yields, respectively; the last figure (j) shows the spectral moments of each case in the $\Delta E_s-T_s$ space comparing with the Maxwellian Locus, from which we can visually distinguish whether the suprathermal distribution of proton is significant or not.

Firstly, as shown in figures \ref{fig4}(a-b), as $\beta$ increases and keeps the initial temperature unchanged, the magnetic energy converted into proton kinetic energy decreases, so its spectral peak shift and effective temperature both decrease, resulting in yield decrease. Roughly inversely proportional to the $\beta$ value, it shares similarities with the estimation of the maximum energy, as shown in (c). Secondly, as the inflow velocity increases, according to the PSA analysis, it can be seen that the spectral center shift will increase significantly (d). When the suprathermal protons are accelerated downstream then re-confined and thermalized by the magnetic field, the downstream proton temperature also increases significantly (e), and generates more alpha particles (f). Then, as the length of the current sheet becomes longer, the acceleration distance and time of the protons become longer, so $\Delta E$(g) and $T$(h) increase significantly, consequently the nuclear reaction yield (i) increases with the scale of the current sheet.

Finally, as shown in figure \ref{fig4}(j), all spectral moments data are plotted against the Maxwellian Locus. Except for the case of S=20, other results clearly show that the proton has a nonthermal energy spectrum, the peak energy is about $75\ \rm{keV}$, and the effective temperature is $30\ \rm{keV}$, which is consistent with the simulation result. Moreover, it can be found that both the increased inflow velocity and the decreased $\beta$ value can make the proton nonthermal spectrum have a larger effective temperature without significant change to the peak energy. Conversely, an increase in the size of the reconnected current sheet significantly increases the peak energy and effective temperature of the proton suprathermal distribution. In addition, the applicable scope of the PSA method is limited by the yields and the spectral moments, so it is estimated from the fitted curves that the applicable parameters of PSA are about: $S>60,\ \beta<64,\ M_s>0.5$.

\subsection{\label{sec:level3-3}Effects from secondary alpha particles and the electromagnetic field}

The above product spectrum analysis is based on an ideal detection environment, where we ignore the distortion of the high-energy alpha particle energy spectrum by the magnetic field and the secondary decay reaction of the pB reaction, $Be^* \rightarrow 2\alpha_1 + 3.03\ \rm{MeV}$. However, the magnetic field deflection of high-energy alpha particles and secondary alpha particles in experiments must be taken into account. Thus more detailed PIC simulations have been carried out to demonstrate that the magnetic field deflection effect and secondary particles have no substantial effect on the above-mentioned analysis approach.

Firstly, we have re-imported the electromagnetic field, particle positions, momentum, and weight information of alpha and beryllium from the reconnection process at $t_0 = 0.4 t_d$ (the moment of the strongest electromagnetic field) into the new simulation cases and expanded the simulation area by a factor of 3. The simulation time has been increased until most of the alpha particles could escape from the electromagnetic field. Secondly, a newly developed two-body decay module is used, which is based on the nuclear reaction module. Because the decay time is much smaller than the simulation step, it is set that the $Be^*$ generated at each step will decay in the next simulation step. It is reasonable to assume that in the COM frame, the decayed secondary $\alpha_1$ particles are isotropically emitted, but are anisotropically converted to the laboratory frame.

\begin{figure*}[htbp]
	\centering
	\includegraphics[width=0.95\textwidth]{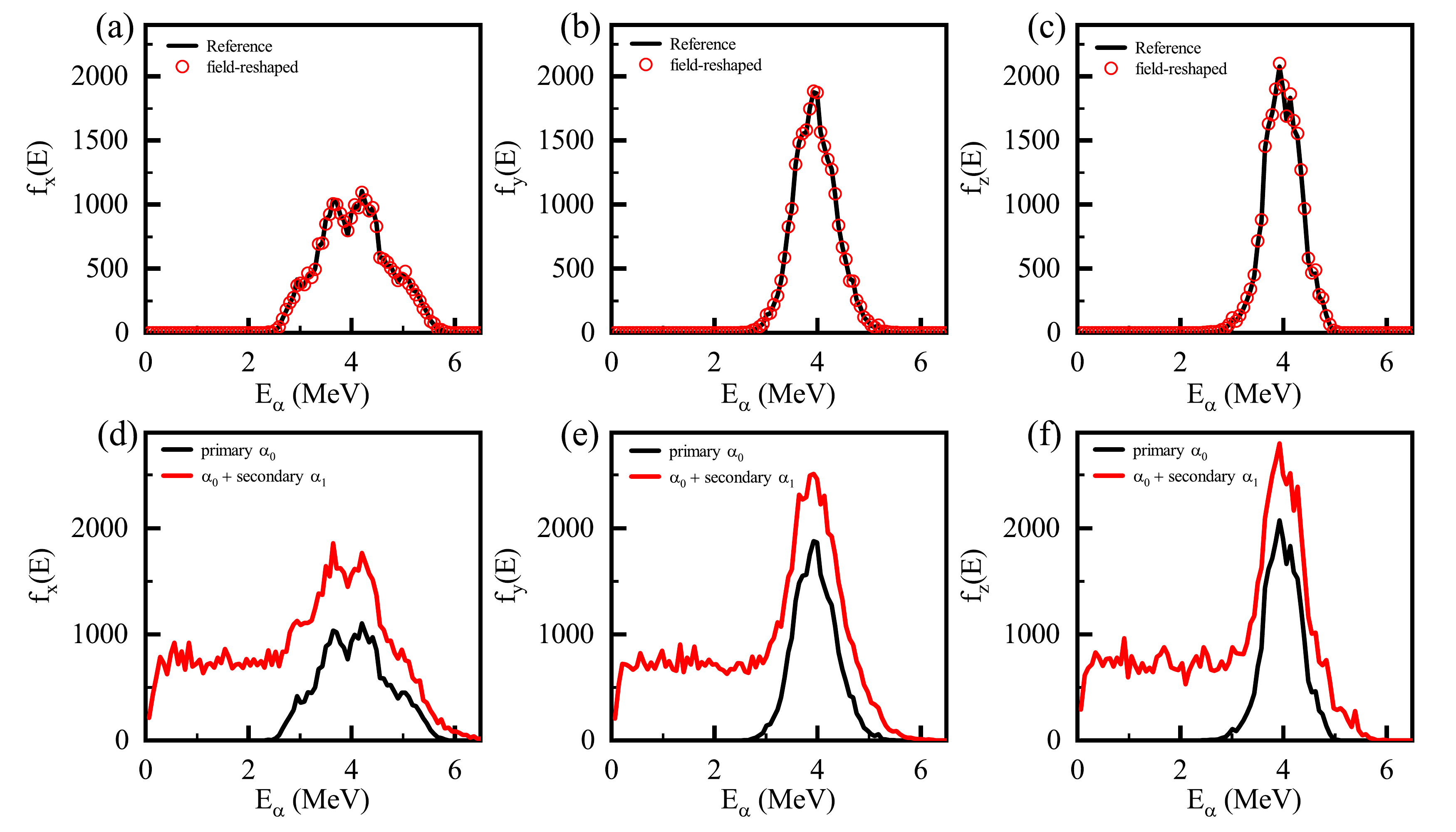}
	\caption{(color). The influence on the alpha energy spectrum from magnetoelectric field and secondary decay reaction. (a), (b) and (c) The comparison between ideal energy spectrum and magnetoelectric field distorted energy spectrum probed in x-, y- and z-direction, respectively. (d), (e) and (f) The comparison between ideal energy spectrum and energy spectrum including secondary alpha particles probed in x-, y- and z-direction, respectively. }
	\label{fig5}
\end{figure*}

Then, we count the product spectrum of the whole alpha population, as shown in figures \ref{fig5}. It is obvious that the magnetic field deflection has a negligible effect on the product spectrum, without modifying the spectral peak and spectral width, in (a-c). Moreover, when the secondary alpha particles are included, the product spectrum shows a plateau structure in the low-energy region and raises the energy spectrum as a whole, but remains the spectrum structure of primary $\alpha$ particles, whose spectral peak and spectral width do not change significantly, as shown in (d-f). Thus, we demonstrate that the deflection of the electromagnetic field and the secondary $\alpha$ particles do not have a significant effect on the spectral moments, and the product spectrum analysis approach is still applicable.

\section{\label{sec:level4}Summary and discussion}

In summary, We propose an approach to obtain information on the ion nonthermal distribution in HED plasma by diagnostic high-energy fusion product spectrum, which is called the product spectrum analysis approach. Firstly, we deduced the directional energy spectrum of the product after the reactant with suprathermal spectrum with directional drift velocity participates in the nuclear reaction and obtained the relationship between the spectral peak, the spectral width of the product spectrum, and the peak energy, effective temperature of the suprathermal spectrum. Then, the correctness of the PSA approach is verified by comparison with the PIC simulation results with the self-consistently expanded nuclear reaction calculation module. Furthermore, taking the laser-driven magnetic reconnection experiment as an example, we have performed a series of simulations and used the PSA approach to diagnose the suprathermal tail spectrum of protons accelerated by magnetic reconnection, and we have obtained results consistent with the simulation. Finally, through the analysis of yield and spectral moments, the applicable parameters range of PSA for magnetic reconnection is estimated as $S>60,\ \beta<64,\ M_s>0.5$.

From the conclusion of this paper, the PSA approach based on nuclear reaction can be applied to these experimental schemes \cite{RN558,RN1299} in the future, because the conditions mentioned above are satisfied to activate a considerable number of fusion products. 

we acknowledge that the product spectrum analysis in this paper is very idealistic, for example, the product spectrum can be distorted by alpha particles being deflected by the electromagnetic field or being stopped by collision with the electron in plasma, and the secondary decay reaction of the pB reaction which can produce indistinguishable alpha particles. Thus, we have discussed the distortion effect of the product spectral moments in the presence of the electromagnetic field and secondary reactions through more detailed simulations. The simulation results show that the deflection of alpha particles caused by the electromagnetic field is tiny enough to be negligible and that the relatively low-energy secondary products only increase the background of the product spectrum without affecting its spectral moments. In addition, although we only deduced the product spectrum produced by the simple nonthermal ion spectrum, in practical applications, we can determine the complexity of the nonthermal ion spectrum by the number of product spectrum peaks, that is, determine $j$ in (\ref{eq1}). Therefore, to sum up, the nuclear diagnostics proposed in this paper with the PSA approach have applicability, accuracy, and universality.

Furthermore, the product spectrum analysis is not only suitable for diagnosing the acceleration effect of reconnection but also has a great potential for other HED plasma experiments that are difficult to detect, such as ICF, laser-driven collisionless shocks, and beam diagnostics of laser-driven ion sources. In addition, because the number of fusion reaction events strongly depends on the reaction cross-section, the nuclear reaction can be appropriately selected to match the characteristic energy region of a particular experiment.

\ack{This work is supported by the National Key Programme for S\&T Research and Development of China, Grant No. 2022YFA1603200; the Science Challenge Project, No. TZ2018005; the National Natural Science Foundation of China, grant No. 12135001, 11921006 and 11825502; the Strategic Priority ResearchProgram of the Chinese Academy of Sciences, grant No.XDA25050900. B.Q. acknowledges support from the National Natural Science Funds for Distinguished Young Scholars, grant No. 11825502. The simulations are carried out on the Tianhe-2 supercomputer at the National SupercomputerCenter in Guangzhou.}

\section*{References}
\bibliography{Ref.bib}

\end{document}